\documentclass[12pt,a4paper]{article}   
\usepackage{latexsym,graphicx,amssymb,amsmath} 
\usepackage{setspace,bm}

\newcommand{\bea}{\begin{eqnarray}}
\newcommand{\eea}{\end{eqnarray}}
\newcommand{\be}{\begin{equation}}
\newcommand{\ee}{\end{equation}}

\newcommand{\rt}[1]{{}}
\newlength{\szovszel}
\newlength{\slashszel}

\begin{document}

\title{}
\title{Axion halo around a binary system of dwarf stars} 
\author{A. Patk\'os\\
Institute of Physics, E\"otv\"os University,\\
 1117 P\'azm\'any P\'eter s\'et\'any 1/A, Budapest, Hungary}

\vfill

\maketitle

\begin{abstract}
 The gravitational field of a clump of ultralight axion like particles (ALPs) in its core with a rotating binary system of dwarf stars is computed. It is established that the induced quadrupole mass moment  of the clump is controlled parametrically by the $M_a/M$ mass ratio of the axion clump and the the binary core.  

\end{abstract}

\section{Introduction}
Fibdibg evidence for the existence of stars composed purely from gravitating light scalar particles represents durable challenge to astrophysical research \cite{kaup68,ruffini69}. Various aspects of stationary equilibrium configurations where gravitational attraction is compensated by the kinetic pressure of the constituents were investigated by many authors \cite{tkachev86,tkachev91,kolb93,kolb94,khlopov98,eby15,braaten16,guth15}.  
Dynamics of the scalar star formation has been explored by kinetic simulation of ensembles of gravitationally interacting free particles 
\cite{levkov18}.

Particularly interesting is the research direction where the galactic halo formed by ultralight constituents is  built from superpositions of quantum waves. In this case the kinetic pressure compensating the gravitational attraction has quantum origin, and it would represent a quantum coherent phenomenon on the largest known scale.  The original proposition \cite{widrow93} has been baptised as $\psi DM$ by Schive {\it et al.} \cite{schive14} emphasizing the role of quantum uncertainty countering gravitation below the Jeans-scale. Applying this balance requirement to dwarf spheroidal galaxies a lower limit for the mass of the superlight dark matter particles was deduced. More recently progress has been achieved in self-consistent determination of the quantum superposition reproducing the observed dark matter halo density profile of dwarf spheroidal galaxies \cite{lin18,yavetz22,zimmermann24}.  

With the advent of black hole observations research has intensified on overdensities of axion like particles (ALP) producing primordial black holes (BH) in an era preceding inflation. Such objects would not evaporate till today if their mass is larger than $10^{-15}M_\odot$. In the gravitational collapse of axions also the emergence of BH pairs has non-zero chance to occur \cite{raidal19}.  Around this kind of binary BH centres of gravitational force the surrounding axion minihalo might further condensate, eventually producing a scalar star\cite{hertzberg20}. 

The equilibrium ALP configuration around a single BH is spherically symmetric. In case of non-relativistic motion of the halo particles the gravitationally bound axion clump forms a so-called gravitational atom. Higher energy configurations with non-zero angular momentum might also arise dynamically.  One scenario considers a second BH falling on a gravitational atom, which resonantly induces transitions to configurations of nonzero quadrupole (and possibly also higher) moments \cite{baumann20,takahashi22}. Such transitions would produce characteristic observable effects in the gravitational waves emitted by the system. 

In this contribution I wish to discuss the interaction of ALPs with another gravitationally bound compact system, binaries of dwarf stars. Systematic search for brown dwarfs has been started in 1990s with observing transiting light curves arising during the passage of brown dwarfs in front of light emitting stars. Very soon binary systems consisting of an ordinary white dwarf star and an accompanying brown dwarf were discovered. About 5-6\% of the known brown dwarfs has a lighting star companion \cite{fontanive18}. From statistical analyses one estimates the separation of the partners in the range of 1.5-1000 au. The mass ratio of the members peaks around unity. The even more difficult observation of a system consisting of two brown dwarfs with $\sim 1$ au separation has been announced very recently \cite{calissendorff23}. The masses of the partners were estimated to lie in the range of 8-20 $M_{Jupiter}$, The period of the rotation lies between 5 to 9 years. The corresponding power of gravitational radiation is $\approx 10^{11}$erg/s by a simple textbook estimate \cite{hartle03}, hopelessly low for present instruments.  More encouraging is  a very spectacular recent report on a rather massive $(M_{BD}\approx 80 M_{Jupiter}) $ brown dwarf transiting in front of a low mass star ($M_*\approx 0.13 M_\odot$). They are very tightly bound with a rotational period of $\sim 2$ hours\cite{el-badry23}. The Keplerian separation is less than the size of our Sun. In this case the simple estimate of the intensity of gravitational radiation gives  nearly 4\% of the electromagnetic radiation power of the Sun. These discoveries motivate us to investigate the structure and dynamical features of ALP clumps around a binary brown dwarf core.

In our analysis presented below the orbiting binary gravitational system will be treated as a pointlike source characterized by the lowest (possibly time-dependent) multipoles of its density distribution. An obvious condition for this is that the Compton wavelength of ALPs should be much larger than the size of the binary core. The latest brown dwarf discoveries offer a realistic ALP mass range for this to be satisfied. The radius of the Sun is $R_\odot\sim 10^6$ km, 1au $\sim 10^8$ km. For an ALP of mass $10^{-n}$ eV the Compton wavelength ($1/m_a$) is at least 100 times larger than the characteristic size of the source in the first case for $n\geq 18$, in the second for $n\geq 20$. This mass range corresponds to the class of {\it ultralight} ALPs. After determining the density distribution produced by the binary source and the gravitational self-interaction of the axionlike particles, one has to check also if the condition that the clump size $R$ exceeds the Compton wavelength of the particle $1/m_a$, e.g. $m_aR>1$ is fulfilled.  

Below we shall determine the profile function of the ALP clump in an approximation, where one truncates the multipole expansion of the gravitational field of the (pointlike) binary core at its quadrupole moment. The particle distribution will be composed from the lowest energy configurations of the $l=0,2$ angular momentum channels. For the gravitational binding energy estimates a variational strategy \cite{guth15} will be applied (see also \cite{schiappacasse18,patkos22,patkos23}). 
The quadrupole deviation of the ALP profile function from spherical symmetry will be determined to linear order. The temporal variation of the elements of the quadrupole tensor of the binary brown dwarf system induces time dependence into the quadrupole piece of the ALP profile. The resulting additional gravitational radiation might offer further insight into the nature of the hypothetical ultralight constituents of matter. 

\section{Determination of the axion halo profile}

Our simplified model for the binary system of two brown dwarfs consists of two $M/2$ mass objects orbiting with angular velocity $\omega$ along a circle of radius $d$ and located in diametrically opposite positions. The gravitational potential will be truncated at quadrupole order
\bea
&\displaystyle
V_N({\bf x})=-\frac{G_NM}{2}\left(\frac{1}{|{\bf x}-{\bf d}|}+\frac{1}{|{\bf x}+{\bf d}|}\right)
\approx-\frac{G_N }{r}\left(M+\frac{1}{r^2}\Theta_{2m}Y_{2m}(\hat{\bf x})\right),\nonumber\\
&\displaystyle
\qquad \Theta_{2m}=\frac{4\pi}{5}Md^2Y_{2, -m}(\hat{\bf d}(t)).,
\label{binary-dwarf-potential}
\eea 
In natural units $(\hbar=c=1)$ the quadrupole moment has inverse mass scaling dimension. Choosing the plane of the orbit for the $(x,y)$-plane, only indices $m=0,2,-2$ contribute to the above sum over $m$. The time dependence of $\bf d$ leads to the time dependence of $\Theta_{2,\pm 2}$. Also one can exploit that $Y_{22}^*=Y_{2,-2}$ and $Y_{2m}(-\hat{\bf d})=Y_{2m}(\hat{\bf d}), m=0,2,-2$. 
 The unit vector $\hat{\bf d}(t)$  points to one of them from the origin, $\hat{\bf x}$ points to the direction of the observation. (The detailed structure of the binary dwarf beyond the data $M, \Theta_{2m}$ does not play any role in the discussion below.)

The energy of the axion "halo" around the binary core is given by
\bea
&\displaystyle
H=\int d^3x\frac{1}{2}\left[\dot a^2({\bf x},t)+(\nabla a({\bf x},t))^2+m_a^2a^2({\bf x},t)\right]\nonumber\\
&\displaystyle
+\int d^2x\rho_a({\bf x},t)V_N({\bf x},t)
-\frac{G_N}{2}\int d^3x\int d^3y \frac{\rho_a({\bf x},t)\rho_a({\bf x},t)}{|{\bf x}-{\bf y}|}.
\eea
In the second line of the above expression the first term gives the energy of particles of mass density $\rho_a$ moving in the gravitational potential $V_N$, while the last term corresponds to the energy of the gravitational attraction among the ALPs constituting the halo.

The assumption for a non-relativistic motion of the particles is reflected in the following parametrisation of the axion field:
\be
a({\bf x},t)=\frac{1}{\sqrt{2m_a}}\left(\psi({\bf x},t)e^{-im_at}+\psi^*({\bf x},t)e^{im_at}\right).
\ee
The slowly varying function $\psi$ is normalized to the number of particles the halo consists of:
\be
\int d^3x|\psi({\bf x},t)|^2=N_a,
\label{normalisation-condition}
\ee
which implies $\rho_a({\bf x},t)=m_a|\psi({\bf x},t)|^2$.
Because of the assumed slow variation of $\psi({\bf x},t)$ only the first time derivative is retained in its equation of motion:
\bea
&\displaystyle
\dot\psi({\bf x},t)=-\frac{1}{2m_a}\triangle\psi({\bf x},t)+V_Nm_a\psi({\bf x},t)
-G_Nm_a^2\int d^3y\frac{|\psi({\bf y},t)|^2}{|{\bf x}-{\bf y}|}\psi({\bf x},t).
\label{sch-equation}
\eea

The quadrupole part of (\ref{binary-dwarf-potential}) induces a piece into the profile funcion $\sim Y_{2m}$. This piece will be determined perturbatively to leading order, therefore it will be proportional also to the dimensionless combination $m_a\Theta_{2m}$ .  In the ansatz chosen for the approximate solution of (\ref{sch-equation}) a coefficient function is introduced in both angular momentum channels depending on the radial coordinate scaled by a characteristic size parameter $R$. 
\bea
&\displaystyle
\psi({\bf x},t)=e^{i\mu t}(\psi_0({\bf x})+\Delta\psi_0({\bf x}))
=e^{i\mu t}w(F_0(\xi)+\tilde F_{2m}(\xi)m_a\Theta_{2m}(d)Y_{2m}(\hat{\bf x})),\nonumber\\
&\displaystyle
 \qquad \xi=\frac{r}{R},\quad \hat{\bf x}=\frac{\bf x}{r}, \quad r=|\bf x|.
\label{profile-ansatz}
\eea
The $R$ parameter characterising the size of the axion clump will be determined variationally. $w$ is a constant to be found from the normalisation (\ref{normalisation-condition}) . 

In the calculation described below one adopts an approximation scheme where the quadrupole piece of the gravitational potential acts perturbatively on the profile of the axion clump relative to the spherically symmetric part of the interaction.  This assumption means that in (\ref{normalisation-condition}) we work to linear order in $\Delta\psi$. Then the normalisation reads as
\be
w^2R^3\left(4\pi\int d\xi \xi^2 F_0^2(\xi)\right)\equiv w^2R^3C_2=N_a.
\ee
Also the radial dependence of the quadrupole part of the profile will be the same for all values of $m$: $\tilde F_{2m}=\tilde F_2$.

Our goal is to compute the additional piece of the gravitational potential of the binary star created by the ALP halo far beyond of its extension. One arrives at  its  expression  by the following sequence of equalities (below $\eta=y/R$):
\bea
&\displaystyle
\Delta V_N({\bf x})=- G_N\int d^3y\frac{\rho_a({\bf y})}{|{\bf y}-{\bf x}|}\nonumber\\
&\displaystyle
\approx-G_Nm_aw^2\int d^3y\frac{1}{|{\bf y}-{\bf x}|}\Bigl[F_0^2(\eta)
+2F_0(\eta)\tilde F_2(\eta)m\Theta_{2m}(d)Y_{2m}(\hat{\bf y})\Bigr]\nonumber\\
&\displaystyle
=-G_Nm_aw^2\Bigl[\frac{1}{r}\int d^3yF_0^2(\eta)
+\frac{8\pi R^5}{5r^3}\int d\eta\eta^4F_0(\eta)\tilde F_2(\eta)m_a\Theta_{2m}Y_{2m}(\hat{\bf r})\Bigr].
\eea
From the very last line one reads off the contribution of the ALP-halo to the quadrupole moment of the system. The complete moment is the sum of this and the original:
\be
\Theta^{sum}_{2m}=\Theta_{2m}\left(1+\frac{8\pi N_a(m_aR)^2}{5C_2}\int d\eta\eta^4F_0(\eta)\tilde F_2(\eta)\right).
\ee
Clearly, the square of the expression in the bracket will multiply the power of the gravitational radiation. Therefore the parametric dependence of $\tilde F_2$ on the dimensionless quantities $N_a, m_aR, G_Nm_a^2$ will be decisive in estimating the effect of the halo on the gravitational power.
 
\section{Determination of $\tilde F_2$}

In this section we determine $\tilde F_2$ which is the $l=2$ admixture to the spherically symmetric profile function $F_0(\xi)$ under the action of the quadrupole part of the gravitational potential. First we write the operator on the right hand side of (\ref{sch-equation}) as a sum:
\bea
&\displaystyle
H=H_0+H_I,\qquad H_0=-\frac{1}{2m_a}\triangle-\frac{G_NMm_a}{r},\nonumber\\
&\displaystyle
H_I=-\frac{G_Nm_a}{r^3}\Theta_{2m}Y_{2m}(\hat{\bf d}(t))-G_Nm_a^2\int d^3y\frac{|\psi(y)|^2}{|{\bf x}-{\bf y}|}.
\eea
The eigenvalue problem of $H_0$ is the gravitational analog of the hydrogen atom of quantum mechanics. The corresponding  eigenvalue-eigenfunction pairs in the $l=0,2$ channels are denoted as $\mu_0, F_0$ and $\mu_2, F_2Y_{2m}$, respectively. (Be careful: the function $F_2$ is not the admixture $\tilde F_2$, we are after!) 

The second term of $H_I$ corrects the value of $\mu_0$ in the first order of perturbation theory. In its evaluation one can neglect in the kernel of the operator the $l=2$ admixture of $\psi_0$ which is perturbatively of higher order. Then the following expression can be readily obtained:
\bea
&\displaystyle
\mu_0N_a=w^2\int d^3x\Biggl[\frac{1}{2m_a}(\nabla F_0(\xi))^2-\frac{G_NMm_a}{|{\bf x}|}F_0^2(\xi)
-G_Nw^2m_a^2\int d^3y\frac{F_0^2(\eta)}{|{\bf x}-{\bf y}|}F_0^2(\xi)\Biggr],
\eea
where the quantities $w,\eta,\xi$ were introduced in the previous section. 
This expression displays a more transparent dependence on the characteristic dimensionless parameter combinations $N_a, m_aR$ and $G_Nm_a^2$, when one writes the integrals in terms of  the scaled variables $\eta,\xi$:
\be
\mu_0 N_a=m_aN_a\frac{1}{C_2}\left[\frac{D_2}{2}\frac{1}{(m_aR)^2}-\frac{G_Nm_a^2}{m_aR}\left(\frac{B_4}{C_2}N_a+\frac{M}{m_a}C_1\right)\right],
\label{grav-eigenvalue}
\ee
where the following integrals of the profile function appear:
\bea
&\displaystyle
C_n=4\pi\int_0^\infty d\xi\xi^nF_0^2(\xi),\qquad D_n=4\pi\int_0^\infty d\xi \xi^nF_0^{\prime 2}(\xi),\nonumber\\
&\displaystyle
B_4=32\pi^2\int_0^\infty d\xi\xi F_0^2(\xi)\int_0^\xi d\eta\eta^2F_0^2(\eta).
\eea

In similar steps one finds the expression of $\mu_2$ with first perturbative order accuracy:
 \be
\mu_2=\frac{m_a}{I_2}\left[\frac{1}{2(m_aR)^2}(K_2+6I_0)-\frac{G_Nm_a^2}{m_aR}\left(\frac{M}{m_a}I_1+\frac{N_a}{C_2}I_{J1}\right)\right],
\ee
with
\bea
&\displaystyle
I_n=\int_0^\infty  d\xi\xi^nF_2^2(\xi),
\qquad
 K_2=\int_0^\infty d\xi\xi^2\left(F_2^{\prime}(\xi)\right)^2,\nonumber\\
&\displaystyle
I_{J1}=4\pi\int_0^\infty d\xi\xi\int_0^\xi d\eta\eta^2
\left[F_2^2(\xi)F_0^2(\eta)+F_2^2(\eta)F_0^2(\xi)\right],
\label{integral-1}
\eea
Here we use the same radial profile function $F_2(\xi)$ for all 5 components of the quadrupole eigenfunction, which is chosen $wF_2(\xi)Y_{2m}(\bf{\hat x})$, for formal uniqueness.

The best estimate for the eigenvalue $\mu_0$ corrected by the nonlinear term of $H_I$ with a conveniently chosen zeroth order profile function $F_0(\xi)$  is found by minimizing the right hand side of (\ref{grav-eigenvalue}) with respect ot $m_aR$ and keeping $N_a, G_Nm_a^2, M/m_a$ fixed \cite{guth15,patkos22,patkos23}. 
The optimal estimates for $m_aR$ and $\mu_0$ are the following: 
\bea
&\displaystyle
(mR)_{opt}=D_2\left[G_Nm_a^2\left(\frac{B_4}{C_2}N_a+C_1\frac{M}{m_a}\right)\right]^{-1},\nonumber\\
&\displaystyle
\mu_{0, opt}=-\frac{m_a}{2C_2D_2}\left[G_Nm_a^2\left(\frac{B_4}{C_2}N_a+C_1\frac{M}{m_a}\right)\right]^2.
\eea
Although in principle one can optimize $\mu_2$ independently, we will be satisfied using the same scale $R$ also for $F_2$.
 
Let us discuss the consistency of the applied approximations  against the parameter range presented in the introduction. Choosing $m_a\sim 10^{-17}$eV one finds with $M\sim M_{Jupiter}$ the following order of magnitude values
\be
G_Nm_a^2\sim 10^{-90},\qquad \frac{M}{m_a}\sim 10^{80}.
\ee
The order of magnitude of the combination of profile function integrals (e.g. $D_2 C_1$) is at most ${\cal O}(10^2)$.  Therefore
\be
(m_aR)_{opt}\sim {\cal O}(10^{-2})10^{10},\qquad |\mu_{0 opt}|\sim 10^{-16}m_a.
\ee
The consistency conditions $m_aR>1$ and $|\mu_0|<<m_a$ are thus fulfilled. The mass contained in the halo around a Jupiter-size brown dwarf binary is well approximated therefore as $N_am_a$. Choosing $N_a$ the same order of magnitude as $M/m_a$ leads to  $M_{halo}\sim M_{Jupiter}$. One can quickly check that the consistency conditions are satisfied even for the high mass ($\sim 10^2M_{Jupiter}$) transiting brown dwarf candidate announced in Ref. \cite{el-badry23}.

The first term of the operator $H_I$ which corresponds to the quadrupole part of the gravitational field of the binary core has nonzero matrix element between $F_0$ and $F_2Y_{2m}$:
\bea
&\displaystyle
\langle l=2,m|H_I|0\rangle=w^2
\int d^3yF_2(\eta)Y_{2m}^*(\hat{\bf y})
\left(-\frac{G_Nm_a}{|\bf{y}|^3}\Theta_{2p}Y_{2p}(\hat{\bf y})\right)F_0(\eta).
\label{quadrupole-piece}
\eea
 Therefore it generates the first order perturbative correction of the lowest energy ALP configuration. The leading quadrupole correction of the profile function $\Delta\psi_0$  is determined using the (familiar from quantum) first order perturbative relation
\bea
&\displaystyle
\Delta \psi_0({\bf x})=\frac{wF_2(\xi)Y_{2m}(\hat{\bf x})}
{\mu_0-\mu_2}\frac{\langle l=2,m|H_I|0\rangle}{\langle l=2,m|l=2,m\rangle}
\nonumber\\
&\displaystyle
=-w\frac{G_Nm_a^2}{(m_aR)^3}\frac{I_{20}^{(-1)}}{I_2}\frac{m_a}{\mu_0-\mu_2}m_a\Theta_{2p}(\hat{\bf d})Y_{2p}(\hat{\bf x})F_2(\xi)
\equiv
w\tilde F_{2m}(\xi)m_a\Theta_{2p}(d)Y_{2p}(\hat{\bf x}).
\eea
with
\be
I_{20}^{(n)}=\int_0^\infty d\eta\eta^nF_2(\eta)F_0(\eta).
\label{integral-2}
\ee

\section{Discussion of the result}

In this contribution we computed the gravitational potential of an axion cloud with its quadrupole distortion induced by a rotating binary dwarf star system in its core. 
\be
\Delta V_N=-\frac{G_NN_am_a}{r}
-\frac{ G_N }{r^3}\Theta_{2p}Y_{2p}\frac{8\pi}{5C_2}\frac{I_{20}^{(-1)}I_{20}^{(4)}}{I_2}\frac{m_a}{\mu_0-\mu_2}\frac{N_aG_Nm_a^2}{m_aR}
\ee
The first term is the contribution of the axion clump to the Newton potential outside the compact object.
Adding the second term to the quadrupole piece of the gravitational field of the core one easily recognizes the "amplification" factor of the quadrupole potential due to the axion halo:
\be
Z_{axion}=1-\frac{8\pi}{5C_2}\frac{I_{20}^{(-1)}I_{20}^{(4)}}{I_2}\frac{m_a}{\mu_0-\mu_2}\frac{N_aG_Nm_a^2}{m_aR}.
\ee
If one would optimize both Schrödinger-like eigenvalues $\mu_0$ and $\mu_2$ one would find  parametrically $\mu_2-\mu_0\sim (m_aR)^{-2}$. The same parametric dependence is suggested by the analogy with the Balmer-formula of the hidrogen atom. Then
one can write parametrically
\be
Z_{axion}=1+{\it const.}\times (m_aR)N_a(G_Nm_a^2)
\ee
which by the optimized expression of $m_aR$ leads to
\be
Z_{axion}=1+{\it const.}\times \frac{N_am_a}{(B_4/C_2)N_am_a+C_1M}.
\ee
One can conclude that the amplification of the quadrupole moment parametrically depends mainly on the  ratio $M_a/M=(N_am_a)/M$. If the mass of the axion clump reaches that of the core then it contributes to the gravitational radiation of the system parametrically the same amount as the core itself. 

In order to present a quantitative estimate for the size of the extra gravitational power originating from the axion halo around the binary brown star system one has to evaluate (\ref{integral-1}) and (\ref{integral-2}) with some well motivated choice of the profile functions $F_0$ and $F_2$.
A physically appealing choice offered by the close formal analogy of the lowest energy configurations of the gravitational "atom" with the $1s$ and $3d$ levels of the hydrogen atom. Then the approach in Refs.\cite{yoo22,yoo23} can be followed choosing for the profile functions the following trial expressions:
\be
F_0(\xi)=Q_0e^{-\xi},\qquad F_2(\xi)=Q_2\xi^2e^{-\xi/3}.
\ee
The arbitrary normalisation coefficients $Q_0,Q_2$ do not appear in any physically meaningful quantity. Straightforward elementary integrations yield explicit values for the coefficients, but do not offer any deeper insight. This excercise is left for the readers. 

\section*{Acknowledgements}
This contribution to the Proceedings of the Bolyai-Gauss-Lobachevsky 2024 conference is a modest way to express solidarity  with all physicists of Ukraine trying to maintain research activities  under the present horrifying war conditions.
This research has been supported by grant K-143460 of NKFIH Science Fund.

\end{document}